\documentclass[12pt]{iopart}
\usepackage{epsfig}

\begin{document}

\title[Particle abundances and spectra]{Particle abundances and spectra
in the hydrodynamical description of
relativistic nuclear collisions with light projectiles}

\author{F.Grassi\dag , Y.Hama\dag , T.Kodama\ddag and O.Socolowski Jr.\dag }

\address{\dag\ Instituto
 de F\'{\i}sica, Universidade de S\~{a}o Paulo, 
C. P. 66318, 05315-970 S\~{a}o Paulo-SP, Brazil}

\address{\ddag\ Instituto de F\'{\i}sica, Universidade Federal do Rio de Janeiro, 
C. P. 68528, 21945-970 Rio de Janeiro-RJ , Brazil}

\begin{abstract}
\noindent  We show that a hydrodynamical model with continuous particle
emission instead of sudden freeze out may explain both the observed strange
particle  and pion abundances and transverse mass spectra for light projectile at SPS energy.
We found that the observed enhancement of pion production corresponds,
within the context of continuous emission, to the maximal entropy production.
\end{abstract}



\maketitle

\section{Introduction}

The main purpose of the ongoing and future heavy ion programs at the high
energy laboratories (CERN, BNL) is to investigate the formation and
properties of hot dense matter, in particular, the phase transition from
hadronic matter to quark gluon plasma (hereafter QGP) predicted by Quantum
Chromodynamics (QCD). Various possible signatures of the appearance of the
QGP have been suggested: entropy increase (due to the release of new degrees
of freedom, namely color), strangeness increase (due to enhanced strange
quark production and faster equilibration), $J/\psi $ suppression (due to
color screening or collision with hard gluons) or enhancement at higher
energy (due to recombination of dissociated $c\bar{c}$ pairs), production of
leptons and photons (emitted from a thermalized QGP and unaffected by strong
interactions), etc. These signals have been studied extensively in
experiments (see for example \cite{ha96}). More recently, the observed high
transverse momentum depletion in Au+Au central collisions at RHIC energies
(jet quenching) and also its absence in d+Au collisions, together with the
system-size dependence of mono-jet formation, have been considered as a
convincing evidence of the formation of QGP at RHIC\cite{Miklos}. On the
other hand, some of the above mentioned signals are purely of
thermodynamical nature and sometimes their significance is not well defined
for finite systems. In such cases, a more careful analysis would be
necessary to clarify the effect of finiteness and dynamical evolution of the
system on these signals\cite{Stoecker}. Some authors suggest that the
incident energy dependence of several quantities should be studied 
 and they claim that the set of incident energy dependences of
particle multiplicity, average transverse momentum and kaon-pion ratio as a
whole indicates the appearance of the mixed phase in central collisions at
SPS energies\cite{MG2}. In fact, if we consider one specific signal of
thermodynamical nature, there are many factors which  may give a similar
response of the claimed signal. For example, it is well known that the
hadronic final state interaction also works to suppress $J/\psi $ and the
system size dependence should be carefully studied to extract the
significance of the observed data. Therefore, it is always important to
study and grasp the hadronic effects on other proposed signals. 

A major problem to trace back any signature unambiguously to a quark gluon
phase is that it is still unknown which theoretical description describes
best high energy nuclear collisions. On one extreme, one might use a
microscopic model. In principle, such approach would provide a faithful
realization of the true physical processes if all the relevant physical
degrees of freedom and corresponding interactions were incorporated
appropriately. However, in practice, this is not possible and some
hypothesis and simplifications are necessarily  introduced. For example, we
may restrict ourselves to pure hadronic degrees of freedom, but is is known
that such  models fail\cite{od98} to reproduce simultaneously strange and
non-strange particle data in nucleon-nucleon collisions and central
nucleus-nucleus collisions at SPS 
energies\footnote{Modifications have been attempted to solve the strangeness problems in
hadronic microscopical models, see however [3]}. Partonic microscopic models
are expected to work at energies higher than SPS\footnote{However see 
\cite{ge98}
and references therein.}. One promising approach to describe the final state
interactions in heavy ion collisions is the UrQMD model\cite{br99}.
However, due to the complexity of the calculation and many uncertainties of
input data such as hadronic cross sections, it is not easy to get a simple
physical insight for the behaviour of the observed quantities.

On the other extreme, one might use a thermal or hydrodynamical model. In
such models, it is assumed that a fireball (region filled with dense
hadronic matter or QGP in local thermal and chemical equilibrium) is formed
in a high energy heavy ion collision and evolves. Hydrodynamical models have
been used successfully to describe various kinds of data from AGS to RHIC
energies. In particular, at SPS, they are able to account for strangeness
data but, in some simplest versions, fail to predict large enough pion
abundances (cf. next section). However such an approach has a great
advantage compared to the microscopic models in the sense that  very few
input data are necessary because of the assumption of the local thermal
equilibrium. Furthermore, we know that this macroscopic description works
quite well suggesting that the local thermal equilibrium in heavy ion
collisions is a reasonable approximation. Thus one might ask whether it is
possible to introduce a correction to the local thermodynamical equilibrium
and its effects on the observed particle abundances. The aim of this paper
is to study this problem. We will argue that, within a very simple and
natural picture of particle emission (continuous emission), the observed
particle abundances and spectra, including strange particles are
consistently reproduced. In addition, we show that the non-equilibrium component
leads to an increase of entropy of the system, hence a substantial pion
increase compared to the usual sudden freezeout mechanism.

\section{Hydrodynamical  description with (standard) freeze
  out emission}

In the standard hydrodynamical models,
one assumes  that particle emission, called in this case
freeze out, occurs on a sharp three-dimensional
surface (defined for example by $T(x,y,z,t) = $ constant). 
Before crossing it, particles have a hydrodynamical behavior,
and after, they free-stream toward the detectors, keeping memory
of the conditions (flow, temperature) of where and when they crossed the
three dimensional surface.
The Cooper-Frye formula 
\cite{co74} gives the invariant momentum distribution
in this case
\begin{equation}
E d^3N/dp^3=\int_{\sigma} d\sigma_{\mu} p^{\mu} f(x,p). \label{CF} 
\end{equation}
$d\sigma_\mu$ is the surface element 4-vector  of the freeze out surface $\sigma$
and $f$ the distribution function of the type of particles considered. 
This is the formula implicitly used in all standard thermal and 
hydrodynamical model
calculations. It can be integrated to get the particle abundance, which depends
 only on the freeze out parameters \footnote{In the past few years, models assuming two different freeze outs, respectively chemical (abundances fixed)
and thermal (shape of spectra fixed) have been frequently used. 
(It is then necessary to modify (\ref{CF}) cf. \cite{fo}.) At RHIC, this view 
is being challenged \cite{flor}.}, e.g.
$T_{f.out}$.

Freeze out parameters can be extracted by analyzing
experimental particle abundances.
This has been done by many groups (for a review see e.g. \cite{so97}). 
The models have some
variations among them in particular some fit $4 \pi$ quantities while others 
consider quantities in fixed rapidity window, each approach having its own 
qualities and draw-backs. For a number of these approaches, it was noted
that
 while they can
reproduce
strange particle abundances, they
underpredict the pion abundance.
This was first noted by  \cite{da92} in a study of
NA35 data 
 and emphasized by
\cite{le93,le95a} in an analysis of the WA85 strange
particle ratios  and EMU05 specific net charge 
$D_q \equiv (N^+-N^-)/(N^++N^-)$ (with $N^+$ and N$^-$, the positive
and negative charge multiplicity respectively).
A similar problem arises with the Pb+Pb data from NA49 \cite{to99,go99}.

Various possible improvements have been suggested so
that these models could yield both the correct strange particle
and pion multiplicities: sequential freeze out \cite{cl93},
hadronic equation of state with excluded volume 
corrections \cite{ri97,ye97,go99}
non-zero pion chemical potential \cite{da92,ye97,to99},
equilibrated plasma 
undergoing
sudden hadronization and immediate decoupling \cite{le93,le95a,le98},  etc.
In this paper we follow a different strategy.
We feel that the assumption  of sudden freeze out on a 3-dimensional surface
is a drastic one; in 
addition it is not sustained by simulations using 
microscopic models \cite{micro}. 
So  we study a different 
particle emission mechanism, continuous emission. 

Before we turn to this, let us note that, 
 to compare particle abundances in the continuous emission and freeze out scenarios,
 we will use a simplified
framework to describe the fluid expansion, 
namely we suppose longitudinal expansion
only and longitudinal boost invariance \cite{bj83}.
This approximation allows to carry out
 some 
 calculations analytically and turns the 
physics involved more transparent.
It is implicit however that this description applies at best to the
midrapidity
region and light projectiles 
(for S+S data, transverse expansion must be small, see e.g. \cite{heinz}).
We should therefore consider midrapidity data  such as S+S data
 from NA35 or WA94, or S+W data from WA85. In fact since we want to consider strange particles and non-strange as well, we will concentrate on NA35 data.  
In this simplified framework,
in the case of a fluid with 
freeze out at a constant temperature and chemical potential, 
the Cooper-Frye formula (\ref{CF}) can
be re-written ignoring transverse expansion as \cite{ru87}
\begin{equation}
\frac{dN}{dy p_{\perp} dp_{\perp}}_{\mid y=0} =\frac{gR^2}{2 \pi} 
\tau_{f.out}(T_{f.out},T_0,\tau_0) m_{\perp} 
\sum_{n=1}^{\infty} (\mp)^{n+1} \exp(\frac{n \mu_{f.out}}{T_{f.out}})
K_1(\frac{n m_{\perp}}{T_{f.out}}) \label{CF_Bj}               
\end{equation}
(The plus sign corresponds to bosons and minus, to fermions.)
It depends  on the conditions at freeze out: $T_{f.out}$ and 
$\mu_{f.out}=\mu_{b\,f.out}B+\mu_{S\,f.out}S$, with B and S the baryon number
and strangeness of the hadron species considered, and 
$\mu_{S\,f.out}(\mu_{b\,f.out},T_{f.out})$ obtained by imposing strangeness
neutrality. 
So the experimental spectra of particles teach us in that case
 what the conditions were at freeze out.

\section{Hydrodynamical description with continuous emission}

The notion that particle emission
 does not necessarily occur on a
three dimensional surface but may be continuous was incorporated in
a hydrodynamical description in
 \cite{gr}.
In this model,
the fluid is assumed to have two components, 
a free part plus an interacting part and 
its distribution function reads
\begin{equation}
f(x,p)=f_{free}(x,p)+f_{int}(x,p).
\end{equation}
$f_{free}$ counts all the particles that 
last scattered earlier at some point and
are at time $x^0$ in $\vec{x}$.
$f_{int}$ describes all the particles that are still 
 interacting (i.e. that will
suffer collisions at time $>x^0$).
The invariant momentum distribution is then
\begin{equation}
E d^3N/dp^3=\int d^4x\,
D_{\mu} [p^{\mu} f_{free}(x,p)]. \label{CE}
\end{equation}
$D_{\mu} [p^{\mu} f_{free}(x,p)]$ is a covariant 
divergence in general coordinates and 
$d^4x$ is the invariant volume element.
 A priori formula (\ref{CE}) is sensitive to the whole fluid history
 and not just to freeze out conditions as in formula (\ref{CF}).

In the simplified framework that we will use 
to compare particle abundances in the continuous emission and freeze out scenarios,
we can approximate the equation of continuous emission (\ref{CE}) as
\cite{gr}
\begin{equation}
\frac{dN}{dy p_{\perp} dp_{\perp}}_{\mid y=0} \sim
\frac{2 g}{(2 \pi)^2} 
\int_{{\cal P}=0.5} d\phi  d \eta   
\frac{m_{\perp} \cosh \eta \tau_F \rho d\rho       
+ p_{\perp} \cos \phi \rho_F \tau d\tau }
{\exp((m_{\perp}\cosh \eta - \mu) /T) \pm 1} 
\label{CE_Bj}
\end{equation}
where
${\cal P}$ is the probability to escape without collision calculated
with a Glauber formula,
$\tau_F$
and $\rho_F$ are  respectively
solution of  
${\cal P}(\tau_F,\rho,\phi,\eta;v_{\perp})=0.5$
 and ${\cal P}(\tau,\rho_F,\phi,\eta;v_{\perp})=0.5$ where $v_{\perp}$ is the 
particle transverse velocity.
In (\ref{CE_Bj}), various $T$ and $\mu=\mu_b B + \mu_S S$ appear
($\mu_S$ is obtained from strangeness neutrality), 
reflecting the whole fluid history,
not just $T_{f.out}$ and $\mu_{b\,f.out}$.
This history is known by solving the hydrodynamical equations of a
hadronic gas with continuous emission; it depends only on the initial
conditions $T_0$ and $\mu_{b0}$ at the initial time $\tau_0$ 
(we use the standard value $\tau_0=1\,fm$). 
Therefore (\ref{CE_Bj}) only depends
on the initial conditions. We expect that heavy particles, due to thermal suppression, will bring information on early times when they are more numerous. 
Fast particles can escape more easily from  dense matter, so they will probe early times, whereas
slow particles will probe late times, when they are finally in diluted matter and make their last collision. Light particles such as the pion should probe the whole fluid history.

\section{Results for continuous emission}

In \cite{grs}, we showed that it is possible to find initial conditions for the
hydrodynamical evolution of the fluid that leads to WA85 strange particle
ratios. In this paper, we want to broader this study. First we want to show that not just ratios, but abundances can be reproduced for a certain set of initial conditions. This is not trivial:  for freeze out,
fitting abundances or ratios is not very different because of
volume cancellation,  
  but not for
 the continuous 
emission which is a process sensitive to the whole fluid history,
  particle mass and particle velocity.
Second, we want to check that for the same initial conditions that reproduce abundances, spectra can be reproduced. 

Abundances can be obtained by integrating (5).
Figure 1 shows the allowed region of initial conditions that lead to
the experimental NA35 $y = 0$ values \cite{NA35}
$\Lambda=1.26 \pm 0.22_{p_\perp > 0.5 GeV}$,
$\overline{\Lambda}= 0.44 \pm 0.16_{p_\perp > 0.5 GeV}$,
$K_S^0= 1.30 \pm 0.22_{p_\perp > 0.62 GeV}$,
$h^-=26 \pm 1$ and $p-\overline{p}= 3.2 \pm 1.0 $.
We do not use the $K^+$ and $K^-$ abundances because they were  measured 
outside the mid-rapidity region.
Considering only strange particles, the allowed window is
$T_0\sim 183-188$ MeV, $\mu_{b0}\sim 70-125$ MeV, with an ideal hadron gas equation of state and 
the strangeness saturation factor (a multiplicative factor for (5))
$\gamma_s=1.3$.
Including the $p-\bar{p}$ abundance decreases the window for $\mu_{b0}$ to
$\sim 70-105$ MeV. Finally including the abundance of negative particles leads
to the small window
$T_0\sim 185$ MeV and $\mu_{b0}\sim 100$ MeV.

Using a more sofisticated equation of state, the value of $T_0$
might be decreased \cite{grs} by some 10-15 \% i.e. to 155-165 MeV,
compatible with  (i.e. below) QCD lattice values for the phase transition
temperature from QGP
to hadronic matter.
Our value of $\gamma_S$ is above 1 and this might look surprising.
However, its value is decreased by some 15 \% when looking at a more realistic
equation of state. In addition, we have imposed  strangeness
neutrality, it is possible that this is a too strong constraint when
analyzing data taken in a very restricted rapidity region (see
\cite{so94} where a similar problem was encountered). Note that
using a larger value
of $\gamma_s$, the size of the allowed window for initial 
conditions in figure 1 increases. Aside of the uncertainty in the equation 
of state,
there are other factors
that influence the precise
location and size of the window but few:  value of the cross section, $=2\, fm^2$, 
(taken constant for simplicity here) in the Glauber formula,
value of the cutoff  ${\cal P}=0.5$ which in fact is equivalent to a change 
in the cross section \cite{gr},
and value of the initial time for
the hydrodynamical evolution for which the canonical value of $\tau_0=1\, fm$
 was assumed.

With the initial conditions determined by figure 1, 
particle spectra can be computed and compared with (all rapidity) NA35
 data.
This is shown in figure 2. The agreement is reasonable. No decays have been included, this should in particular lead to an improvement of the low $m_\perp$
pion spectrum.

Therefore there exist
 initial conditions of the hydrodynamical expansion
such that NA35 strange and non-strange particle abundances and spectra
can be reproduced
simultaneously without extra assumption,
in contrast to  some of the freeze out models mentioned above.
In fact it is puzzling that the continuous model leads  to so many more pions than the freeze out
model and it is necessary to investigate the reason.

To  illustrate more precisely the difference between (simple) freeze out and continuous 
emission 
scenarios,
 we compare in table 1 results from both scenarios, with
 $T_0=185 MeV$ and
$\mu_{b0}=100 MeV$ for the continuous emission case,
$T_{f.out}=185 MeV$ 
and
$\mu_{b\,f.out}=100 MeV$ for the freeze out case
(with the scaling factor $\tau_{f.out}$ in (2) taken equal to $ 1\,fm$)
\footnote{For both models, the values chosen for the parameters are typical.}.
For both cases,
 we assume that initially the
matter is a hadron gas
and 
$\gamma_s=1.3$. 
Heavy particles,
in the continuous
emission case, due to thermal suppression,  are mostly emitted
early \cite{grs}, i.e. in similar conditions than in
 the freeze out model in the case $\mu_{b\,0}=0$ and for some longer
 time
if $\mu_{b\,0}\neq 0$. 
Pions in
the freeze out case are too few as  discussed previously: 15.7 instead of $26 \pm 1$.
 Pions in
the continuous emission case, on
the other side, are emitted early and then on, and we get substantially
more of them, 27 in agreement with data\footnote{To compute the number of 
pions in the continuous emission case, the effect of continuous emission on the hydrodynamical  evolution of the fluid is taken into account as in 
 \cite{gr}.}. 

In a hydrodynamical model without shocks and dissipation, entropy is 
conserved. In the usual freeze out scenario, this is an important
point because the initial entropy can be determined from the final 
multiplicity.
To illustrate this connection, let us consider a thermalized massless pion 
fluid. 
In this case, the entropy density is related to the pion density at all temperature by $s=3.6 n_{\pi}$.
Therefore knowing the pion number at freeze out (from data) 
$N_{\pi}^{f.out}$, 
one can infer
the entropy at freeze out $S^{f.out}$ and 
  the initial value of the entropy
\begin{equation}
S_0=S^{f.out}=3.6 N_{\pi}^{f.out}
\end{equation}
 (since  entropy is conserved).

In the continuous emission case,
we can compute the number of free particles in principle from  (4)
\begin{eqnarray}
N_{\pi}^{free} & = & \int d^4x \int d^3p D_{\mu}(\frac{p^{\mu}}{E} f_{free}) \\
        &                   = & \int d^4x D_{\mu} \left[ \frac{g}{(2 \pi)^3}
\int d^3p \frac{p^{\mu}}{E} \frac{1}{e^{(u.p)/T}-1}\frac{\cal P}{1-{\cal
        P}}
\right]\nonumber \\
        &                   = &  \int d^4x a(x) n_{th}^{m=0}(x) \nonumber 
\end{eqnarray}
where
 we used the factorization property arising from
the zero mass assumption \cite{gr}) and introduced a kind of weight,
$a(x)\equiv D_{\mu}(A^{\mu} 
n_{th}^{m=0})/n_{th}^{m=0}$
with 
$A^{\mu}= ( \int d\Omega/(4 \pi)\, {\cal P}/  
(1-{\cal P}),  \int d\Omega \sin \theta \cos \phi /(4 \pi)\, {\cal P}/  
(1-{\cal P}),  \int d\Omega \sin \theta \sin \phi /(4 \pi)\, {\cal P}/  
(1-{\cal P}), \int d\Omega \cos \theta / (4 \pi)\,
 {\cal P}/ (1-{\cal P}))$.

We re-write (7) by assuming that the fluid is divided in small volumes 
$V_{\alpha}(\tau)$ moving with it and time is  discretized with
$\tau_{i+1}=\tau_{i}+\Delta \tau$, so
\begin{equation}
N_{\pi}^{free}= \sum_{\alpha=1}^{\infty}
        \sum_{i=1}^{\infty} \Delta \tau \Delta V_{\alpha}(\tau_i) 
a(\tau_i)  n_{th}^{m=0}(\tau_i)
\end{equation}

At time $\tau_i$,  $\Delta V_{\alpha}(\tau_i)$
contains a mixture of free plus thermalized particles and we suppose that
 the thermalized pions break into 
those which have just done their last collisions (joining the free component
of the fluid)
and those which remain 
interacting. This interacting component is supposed to have reached thermal
equilibrium at $\tau_{i+1}$ and the process of separation between free and 
interacting part repeats itself. We note that
both the separation (free/interacting) process and the re-thermalization one
 (even if incomplete)
 lead to entropy 
increase, which can manifests itself as pions.
Exactly how much entropy is created and how much appears as pions is 
model-dependent, but this effect should always be present. 
Depending on the process involved, this can be a substantial effect. Indeed
as shown in table 1 and discussed above, in the particular case of
 continuous emission with $f_{int} \sim f_{th}$ (complete 
rethermalization), 
the amount of extra pions can even reach 70\%.
It would be worth checking if this is the case with the improved 
description of continuous emission \cite{yh}. 
 The hypothesis $f_{int} \sim f_{th}$  corresponds to the upper
limit of
entropy production. 
It is interesting to note that the observed pion abundance is reproduced just by the maximal entropy production within the present mechanism.
Comparing results from UrQMD and the thermal 
statistical model \cite{br99}, it was shown that the entropy density per baryon density 
increases 
by 15 \%
with increasing times, 
or alternatively  for temperatures decreasing from 161 to 126 MeV, in
a central cell $5 \times 5 \times 5 fm^3$ for Pb+Pb at SPS. 
In our case, for such a central cell, away from the border, we have
initially little continuous emission and the entropy per baryon increases 
slowly. The large increase of pion abundance obtained above occurs in the space-time domain where continuous emission becomes considerable.

\section{Conclusion} 
 
In this paper, we discussed data on strange and non-strange particles
at ultrarelativistic energy with light projectiles,
 from a hydrodynamical point of view. 
Some versions of the standard  model with sudden freeze out can reproduce the strange
particle data but underpredicts the pion abundance, if no extra
assumption is made. In addition, it is usually necessary to assume two different freezeouts, chemical and thermal, to account for both strange particle
abundances and particle transverse mass spectra.
We showed that a hydrodynamical model with a more precise emission
process, continuous emission, can reproduce both the strange and
non-strange particle abundances without extra assumption in addition to being
consistent with other types of experimental data such as transverse
mass spectra. In the two freeze outs case, typical parameters are 
chemical freeze out temperature, baryonic potential and strangeness saturation factor
as well as thermal freeze out temperature and baryonic potential; they are fixed by data.
In the continuous emission, the parameters are: the initial
conditions $T_0$, $\mu_{b0}$ and $\gamma_s$, which are fixed by data;
the average interacting cross section and initial time were chosen to have the canonical values $2\,fm^2$ and $1\,fm$.
We note that while in freeze out models, data give information only on the freeze out conditions, in continuous emission, observables depend on the whole fluid history. Therefore what observables teach us depend on the emission model. 
This point is reinforced by a comparison
of Bose-Einstein correlations for freeze out and continuous emission \cite{hbt}

Our main point is the following: 
in the usual freeze out scenario,
a large pion number may be 
associated with a large entropy (cf. (6)). 
Here we
showed that a large pion number can be generated by continuous emission.
The reason for this increase is entropy generation during the separation and rethermalization processes occurring during continuous emission.
  We stressed that how large is this increase is model-dependent, but in any case this possibility
 sheds a new light on the problem of pion emission at SPS.
For freeze out,
a large experimental value of $N_{\pi}$ implies
 a large
initial entropy $S_0$ and may be considered
a hint of  QGP formation (see e.g. \cite{le93,le95a}).
For continuous emission, a large pion number and entropy may be a natural 
outcome independently of QGP formation.
Consequently,
a better understanding of
particle emission in the hydrodynamical regime
is  necessary to assess the possibility of QGP formation in
relativistic heavy ion collisions.

\section*{Acknowledgements}

This work was partially supported by CAPES, CNPq, FAPERJ and
 FAPESP (2000/04422-7, 2000/05769-0, 2001/09861-1).

\newpage
\begin{table}
\caption{ Comparison of experimental particle abundances with
continuous emission and freeze out predictions for S+S collisions at 
midrapidity.}
\begin{indented}
\item[]\begin{tabular}{|c|l|l|l|} \hline
  & experimental value & continuous emission & freeze out\\  \hline
$\Lambda$ & 1.26$\pm$0.22 & 0.96 & 0.92 \\
$\bar{\Lambda}$ & 0.44$\pm$0.16 & 0.29 & 0.46 \\
$p-\bar{p}$  & 3.2$\pm$1.0 & 3.12 & 1.32 \\
$h^-$     & 26$\pm$1 & 27 & 15.7 \\
$K^0_S$   & 1.3$\pm$0.22  & 1.23 & 1.06\\ \hline
\end{tabular}
\end{indented}
\end{table}
\newpage

\begin{figure}
\begin{center}
\epsfig{file=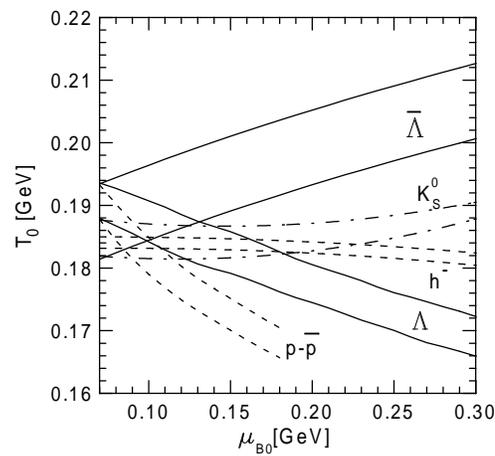,height=6.cm,angle=0}
\end{center}
\caption{Allowed region for the initial conditions determined from the NA35 S+S midrapidity  data.}
\end{figure}

\begin{figure}
\begin{center}
\epsfig{file=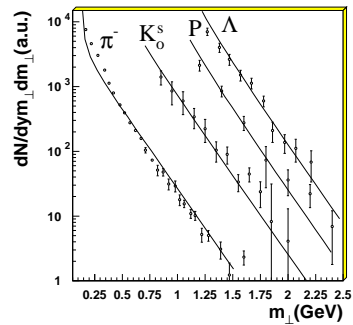,height=6.cm,angle=0}
\end{center}
\caption{Transverse mass spectrum computed for the initial conditions
obtained in figure 1 and comparison with NA35 (all-y) data. No decays included.}
\end{figure}

\end{document}